
\font\twelverm=amr10 scaled 1200    \font\twelvei=ammi10 scaled 1200
\font\twelvesy=amsy10 scaled 1200   \font\twelveex=amex10 scaled 1200
\font\twelvebf=ambx10 scaled 1200   \font\twelvesl=amsl10 scaled 1200

\font\twelvett=amtt10 scaled 1200   \font\twelveit=amti10 scaled 1200
\font\twelvesc=amcsc10 scaled 1200  
\skewchar\twelvei='177   \skewchar\twelvesy='60
\def\twelvepoint{\normalbaselineskip=12.4pt plus 0.1pt minus 0.1pt
  \abovedisplayskip 12.4pt plus 3pt minus 9pt
  \belowdisplayskip 12.4pt plus 3pt minus 9pt
  \abovedisplayshortskip 0pt plus 3pt
  \belowdisplayshortskip 7.2pt plus 3pt minus 4pt
  \smallskipamount=3.6pt plus1.2pt minus1.2pt
  \medskipamount=7.2pt plus2.4pt minus2.4pt
  \bigskipamount=14.4pt plus4.8pt minus4.8pt
  \def\rm{\fam0\twelverm}          \def\it{\fam\itfam\twelveit}%
  \def\sl{\fam\slfam\twelvesl}     \def\bf{\fam\bffam\twelvebf}%
  \def\mit{\fam 1}                 \def\cal{\fam 2}%
  \def\sc{\twelvesc}               \def\tt{\twelvett}
  \def\sf{\twelvesf}
  \textfont0=\twelverm   \scriptfont0=\tenrm   \scriptscriptfont0=\sevenrm
  \textfont1=\twelvei    \scriptfont1=\teni    \scriptscriptfont1=\seveni
  \textfont2=\twelvesy   \scriptfont2=\tensy   \scriptscriptfont2=\sevensy
  \textfont3=\twelveex   \scriptfont3=\twelveex  \scriptscriptfont3=\twelveex
   \textfont\itfam=\twelveit
  \textfont\slfam=\twelvesl
  \textfont\bffam=\twelvebf \scriptfont\bffam=\tenbf
  \scriptscriptfont\bffam=\sevenbf
  \normalbaselines\rm}
\def\journal#1 #2 #3 #4.{{\it #1} {\bf #2}, #3 (#4).}
 \twelvepoint
\baselineskip=20pt
\parindent 20pt
\settabs 4 \columns
\parskip=10pt
\hfuzz=2pt   
\def\line{\hbox to \hsize}
\def\frac #1#2{{#1\over #2}}
\def\Tr{{\rm  Tr\,}}
\def\tr{{\rm  tr\,}}
\def\Psid{\Psi^{\dagger}}
\def\psid{\psi^{\dagger}}

\def\Z{{\cal Z}}

\def\Det{{\rm Det\,}}


\line{\hfil ILL-(CM)-94-22}
\line{\hfil September 1994}
\vskip 1cm
\centerline{\bf DYNAMICS OF $T=0$ BCS}
\centerline{\bf  CONDENSATES}
\vskip 1cm
\centerline{Michael Stone}
\vskip .5cm
\centerline{\it University of Illinois at Urbana Champaign}
\centerline{\it Department  of Physics}
\centerline{\it 1110 W. Green St.}
\centerline{\it Urbana, IL 61801}
\centerline{\it USA}
\vskip 1cm

\line{\bf Abstract\hfil}

Fermi-surface bosonization is used to show that the long-wavelength,
$T=0$,  dynamics of a BCS superfluid or superconductor is described by
a galilean invariant non-linear time-dependent Schr{\"o}dinger
equation. This equation is  of same form as the  Gross-Pitaevskii
equation for a Bose superfluid, but the  ``wavefunction''
 is {\it not} the superfluid order parameter.

\vfil\eject

\line{\bf Introduction\hfil}

At zero temperature the dynamics of a Bose superfluid can be  modeled,
at least qualitatively, by the Gross-Pitaevskii time-dependent
non-linear Schr{\"o}dinger equation [1,2].  It is reasonable to ask
whether the same simple equation may be used as a guide to the zero
temperature condensate dynamics  of a  Fermi system   where the
superfluidity is a result of  $S$-wave BCS pairing. Such
Schr{\"o}dinger dynamics was assumed, for example,  by Feynman in his
discussions of superconductivity [3].  Surprisingly,   no explicit
statement as to the validity of such an assumption is to be found in
the standard superconductivity literature. Indeed a cursory reading of
the classic papers such as refs. [4,5] gives the impression that the
phase mode of the superconducting order parameter obeys a wave equation
with a second-order time derivative, as opposed to the first-order time
derivative found in the time-dependent Schr{\"o}dinger equation. This
impresion is also given by  more recent work [6,7] where the crossover
between weak-coupling BCS and the Bose condensation of preformed pairs.
is discussed.  Recently, however, the case for galilean invariant
Schr{\"o}dinger dynamics  has been made by Ao, Thouless,and Zhu [8] who
were motivated by earlier work on  Berry's phase and vortex motion by Ao,
and  Thouless[9].

First- versus second-order time derivatives  imply very different
motion for vortex defects in the order parameter field. With
second-order time derivatives the members of a pair of initially
stationary vortices accelerate toward each other and annihilate [10].
With Schr{\"o}dinger  dynamics the vortex  and anti-vortex maintain
their separation and drift in parallel  at a steady speed [11], each
being carried  by the superflow induced by the other. A
Schr{\"o}dinger  vortex can only  move relative to the  superflow in
the presence of an external force which balances the Magnus lift on the
vortex core.  (When the superfluid is charged,  the core vorticity is
screened by the magnetic field-induced anti-vortex. The force on a thin
tube surrounding the vortex core is the Magnus lift. For a tube larger
than a magnetic penetration length the Magnus lift is replaced by an
equivalent Lorentz body-force[12].)

 This present  paper is intended to provide an elementary derivation of
 the Gross-Pitaevskii Schr{\"o}dinger equation dynamics from the
 microscopic  $T=0$ BCS model. To do this we will use a simple form of
Fermi-surface bosonization.  In the second section we will briefly
review the standard field-theory formulation of the BCS model. In the
third  and fourth sections we will explain the bosonization method for
calculating the effective action.

Some of the results of this paper will appear in [13].

 \line{\bf 2) Zero Temperature Effective Action for BCS
Superconductors\hfil}

We begin with a quick review of the field-theory approach to
neutral BCS superfluids.

The  partition function of a gas of non-relativistic
spin$-\frac 12$ fermions may be written as a Grassmann path integral
$$
\eqalign{
{\Z}=&\Tr(e^{-\beta H}) \cr
=&\int d[\psi]d[\psid]\exp -\int_0^\beta d^3x d\tau \left\{
\sum_{\alpha=1}^{2} \psid_\alpha(\partial_\tau-\frac
1{2m}\nabla^2-\mu)\psi_\alpha -
g\psid_1\psid_2\psi_2\psi_1\right\}.
\cr}
\eqno (2.1)
$$
The indices $\alpha= 1,2$ refer to the two components of spin. As is usual in
the Matsubara formalism, the Grassmann-valued Fermi fields are
to be taken  antiperiodic under the shift $\tau \to \tau+\beta$.

The effective action we seek to derive will not depend on the details
of  the interaction that gives rise to superconductivity, so, for
simplicity, we have included only a short-range interaction,
$g\psid_1\psid_2\psi_2\psi_1$, in (2.1).  A positive value for $g$
corresponds to an attractive potential. Given an attractive interaction, and
a low enough temperature, the system should be unstable  with respect
to the onset of superconductivity. To detect this instability we
introduce an ancillary  complex scalar field $\Delta$ which will become
the superconducting order parameter. We use it to decouple the
interaction
$$
\eqalign{
{\Z}=\int d[\psi]d[\psid]d[\Delta]d[\Delta^*]
\exp -&\int_0^\beta d^3x d\tau \biggl\{
\sum_{\alpha=1}^{2} \psid_\alpha(\partial_\tau-\frac
1{2m}\nabla^2-\mu)\psi_\alpha  \cr
 -&\Delta^*\psi_2\psi_1-\Delta\psid_1\psid_2
+\frac {1}{g}|\Delta|^2\biggr \}.
\cr}
\eqno (2.2)
$$
The equation of motion for $\Delta$ shows us that $\Delta\equiv g
\psi_2\psi_1$.

We may now integrate out the fermions to find an effective action
for the $\Delta$ field. From this point on we will set the
temperature, $\beta^{-1}$, to zero.

Taking note of the anticommutativity of the  Grassmann fields,
the quadratic form in the exponent can be arranged as a
matrix
$$
S=\int d^3xd\tau (\psid_1,\psi_2)\left
(\matrix{\partial_\tau-\frac{\nabla^2}{2m}-\mu & \Delta \cr
          \Delta^* & \partial_\tau+\frac{\nabla^2}{2m}+\mu\cr}\right)
\left(\matrix{\psi_1\cr\psid_2\cr}\right).
\eqno (2.3)
$$
The fermion contribution to the effective action is  the
logarithm of the  Fredholm determinant of this matrix of differential
operators
$$
S_F=-\ln\Det \left(\matrix{\partial_\tau-\frac{\nabla^2}{2m}-\mu & \Delta \cr
          \Delta^* & \partial_\tau+\frac{\nabla^2}{2m}+\mu\cr}\right).
\eqno (2.4)
$$

We begin by first assuming that $\Delta$ is a constant. Under these
circumstances $S_F$ is given by
$$
S_F=\int d^3x d\tau\left\{
-\int \frac {d^3k}{(2\pi)^3}\frac{d\omega}{2\pi}
\tr\ln \left(\matrix{i\omega+\frac{k^2}{2m}-\mu & \Delta \cr
          \Delta^* & i\omega-\frac{k^2}{2m}+\mu\cr}\right)\right\}.
\eqno (2.5)
$$
It is convenient to introduce the notation $\epsilon=k^2/2m-\mu$,
whence the momentum integral becomes
$$
I=\int \frac {d^3k}{(2\pi)^3}\frac{d\omega}{2\pi}
\ln(\omega^2+|\Delta|^2+\epsilon^2).
\eqno (2.6)
$$
Since everything is spherically symmetric
we can replace the integration over $k$ by an integration over $\epsilon$
$$
I=\int \rho(\epsilon)d\epsilon \frac{d\omega}{2\pi}
\ln(\omega^2+|\Delta|^2+\epsilon^2),
\eqno (2.7)
$$
at the expense of introducing the density of states
$\rho(\epsilon)$.  We evaluate (2.7) by first differentiating with respect to
$\Delta$
$$
\frac{dI}{d\Delta}=\int d\epsilon\frac{d\omega}{2\pi}  \rho(0)
\frac {\Delta^*}{ \omega^2+|\Delta|^2+\epsilon^2}.
\eqno (2.8)
$$
In (2.8) we have approximated $\rho(\epsilon)$ by $\rho(0)$. This
approximation is reasonable  because the integrand is peaked near the
Fermi surface $\epsilon=0$. To obtain a finite value  for the  integral
we must introduce a cut-off.  In phonon-mediated  BCS superconductivity
the interaction naturally dies out at the energy $\epsilon_D$
corresponding to the Debye frequency.  Taking this value for the
cut-off we find
$$
\frac{dI}{d\Delta}=\frac
{\rho(0)}{2}\int_{-\epsilon_D}^{\epsilon_D} \frac
{\Delta^*}{\sqrt{\epsilon^2+|\Delta|^2}}  =-\frac
{\rho(0)}{2}\Delta^*\ln \frac{|\Delta|}{\epsilon_D}.
\eqno (2.9)
$$

Putting this together with the $|\Delta|^2$ part of the exponent,
we find that the effective potential (the action per unit volume
space-time) for $\Delta$ is minimized when
$$
\frac{dV_{eff}}{d\Delta}=\left( \frac
{\Delta^*}{g}+\frac{\rho(0)}{2}
\Delta^*\ln\frac{|\Delta|}{\epsilon_D}\right)=0,
\eqno (2.10)
$$
or  when
$$
|\Delta|=|\Delta_0|= \epsilon_D\exp\left\{-\frac
{2}{g\rho(0)}\right\}.
\eqno (2.11)
$$

The effective potential itself is
$$
V_{eff}=\frac 1{g}|\Delta|^2+\frac
{\rho(0)}{4}\left\{|\Delta|^2\ln\frac{|\Delta|^2}{\epsilon_D^2}-
{|\Delta|^2}\right\}.
\eqno (2.12)
$$

We now wish to investigate the terms in the effective action
$S_F(\Delta)$ that involve space-time gradients of $\Delta$.  Since
$\Delta$ plays the role of a mass-gap we would expect to able to find
an expansion in increasing orders and powers of derivatives of
$\Delta$, each extra derivative being accompanied by a factor of
$|\Delta_0|^{-2}$. A diagrammatic evaluation of the terms in such an
expansion is possible  [14], but we can get physical insight into
the problem  by using some simple tricks, including bosonization.

To elucidate the kind of terms we  can expect to find in the gradient
expansion we begin by considering the effect of uniform twists.
Suppose that the phase of the order parameter varies linearly with
position ,{\it i.e} $\Delta(x)=e^{2ik_sx}\Delta_0$. This should
correspond to a uniform superflow with  velocity $v_s=k_s/m$.

We need to find the Fredholm determinant of
$$
K= \left(\matrix{\partial_\tau-\frac{\nabla^2}{2m}-\mu & \Delta_0e^{2ik_sx} \cr
      \Delta_0^*e^{-2ik_sx} &\partial_\tau+
\frac{\nabla^2}{2m}+\mu\cr}\right).
\eqno (2.13)
$$
Now if
$$
U=\left(\matrix{e^{ik_sx}& 0\cr
                   0    & e^{-ik_sx}\cr}\right),
\eqno (2.14)
$$
then
$$
U^{-1}KU=
\left(\matrix{\partial_\tau-\frac{(\nabla+ik_s)^2}{2m}
                                               -\mu & \Delta_0 \cr
\Delta_0^* &\partial_\tau+\frac{(\nabla-ik_s)^2}{2m}+\mu\cr}\right).
\eqno (2.15)
$$
The determinant is not affected by such a unitary
transformation\footnote{*}{This is not  true for
relativistic systems where the determinant {\it is\/} altered by such
chiral transformations because of anomalies. There are no anomalies
for non-relativistic systems.} and we can now evaluate $\ln \Det K$ by fourier
transforming.
$$
\ln \Det K= \int \frac {d^3k}{(2\pi)^3}\frac{d\omega}{2\pi}
\tr\ln \left(\matrix{i\omega+\frac{(k+k_s)^2}{2m}-\mu & \Delta_0 \cr
          \Delta_0^* & i\omega-\frac{(k-k_s)^2}{2m}+\mu\cr}\right)
$$
$$
=\int \frac {d^3k}{(2\pi)^3}\frac{d\omega}{2\pi}
\tr\ln
\left(\matrix{(i\omega+v_sk)+\frac{k^2}{2m}-(\mu-\frac{k_s^2}{2m})
& \Delta_0 \cr
          \Delta_0^* & (i\omega+v_sk)-
\frac{k^2}{2m}+(\mu-\frac{k_s^2}{2m})\cr}\right).
\eqno (2.16)
$$
The net effect has been to shift $i\omega\to i\omega+v_s k$
and $\mu \to \mu-k_s^2/2m$.

If we  momentarily forget about the shift in the $\omega$
variable, we find that, for small $k$,
$$
S_F(\Delta_0 e^{2ik_sx})=S_F(\Delta_0)-\frac {k^2}{2m}
\frac{\partial S_F}{\partial \mu}.
\eqno(2.17)
$$
Now  $-{\partial S_F}/{\partial \mu}$ is the number density of
the system, so
$$
S_F(\Delta_0 e^{2ik_sx})=S_F(\Delta_0)+\rho\frac {k^2}{2m}.
\eqno(2.18)
$$
This is what we would expect for a uniform flow of the entire
fluid with $v_s=k/m$.

Let us now ask for the consequences of the sideways translation of the
$\omega$ contour. In evaluating the density in (2.18) it is  the
$\omega$ contour integral that determines the occupation numbers of the
various quasiparticle modes. These modes manifest themselves as poles
in the $\langle\psid\psi\rangle$ Green function at
$\omega=\pm\sqrt{\epsilon^2+|\Delta|^2}$, and the contour integral is
to be closed  in such a manner that the negative energy states  are
encircled, and thus counted as occupied.  These negative energy states
are particle-like for $k \ll k_f$ and hole-like for $k \gg k_f$. If
$v_s$ is sufficiently large, poles that were within the contour  before the
$iv_s\cdot k$ shift may no longer be enclosed, and conversely,
previously unoccupied states may be occupied. The physical reason is
that, when seen from the rest frame, the quasiparticle energies in the
moving fluid are doppler-shifted from $\omega(k)$ to $\omega(k) + v_s
k$. The use of a chemical potential $\mu$ to determine
the average number of particles  implies equilibrium with the stationary
walls of the container, and it is the negative energy states as seen
from the container frame that are occupied.  There is a range of $v_s$ for
which the occupation numbers are unchanged, and consequently a critical
value of $v_s$ below which no ``normal'' fluid will be created. Above
$v_{crit}$ we need to replace the $\rho$ in (2.18) by an effective
$\rho_s$. Since the  alteration of occupation numbers tends to reduce
the momentum in the system we have $\rho_s< \rho$. Only the superfluid
fraction is flowing. The normal component stays at rest.

We now consider the  time variation $\Delta=e^{2i\Omega
t}\Delta_0$. This shifts  $\mu\to \mu -i\Omega$   so, for
small $\Omega$,
$$
S_F(e^{2i\Omega t}\Delta_0)=S_F(\Delta_0)+i\rho \Omega.
\eqno (2.19)
$$
Thus we know that  the euclidean action $S_F$ will contain the terms
$$
S_F(\Delta)\approx V_{eff}(\Delta)+i{\rho}\partial_\tau\phi/2+
\frac{\rho}{2m}(\partial_x\phi/2)^2,
\eqno (2.20)
$$
where $\Delta=e^{i\phi(x,t)}|\Delta|$.

Notice that if we work in real  time (as opposed to the Matsubara
imaginary time that we have been using)   then, at least for
$v_s<v_{crit}$, $S_F$ is  invariant  under the combined phase
transformation $\Delta\to (e^{imv_sx-i\frac 12 mv_s^2t})^2\Delta$.
This corresponds to   a simultaneous galilean transformation
acting on all the particles in the fluid.  When the fluid and its
container are moved together the occupation numbers should
remain unchanged even if   $v_s>v_{crit}$. In this circumstance
the contour should be arranged so that the $v_sk$
translation has no effect on the occupation numbers.

\line {\bf 3.) Bosonization\hfil}

We will now use  three-dimensional Fermi-surface bosonization to
directly calculate the  effective action  for slowly varying
$\Delta(x,t)$.  The earliest discussion of higher dimensional
bosonization to have seems to have been by Luther [15]. More recently
the technique has been pursued by Haldane [16] and others [17,18]
interested in the possible breakdown of Fermi-liquid theory in two
dimensional strongly correlated systems [19].
 The  discussion in this section is based on methods introduced in [21].

We will warm up  with a one-dimensional system. We begin by decomposing
the Fermi field into pieces that live near the two fermi points
$$
\eqalign{
\psi_i=&e^{ik_fx}\psi_{i,R}+e^{-ik_fx}\psi_{i,L}\cr
\psid_i=&e^{-ik_fx}\psid_{i,R}+e^{ik_fx}\psid_{i,L}.
\cr}
\eqno (3.1)
$$
The fields $\psi_{R,L}$ contain those modes lying within some  energy  range of
width $\Lambda$
 about  the Fermi points. We are going to assume that the processes we
are interested in affect states only within this energy shell.

We next introduce two two-component fields
$$
\Psi_1=\left(\matrix {\psi_{1,R}\cr\psid_{2,L}}\right)\quad
\Psid_2=\left(\matrix {\psi_{2,R}\cr\psid_{1,L}}\right),
\eqno (3.2)
$$
and their hermitian conjugates.
In terms of (3.2) the fermion  current and  number become
$$
j=v_{f}\sum_{i=1,2}(\psid_{i,R}\psi_{i,R}-\psid_{i,L}\psi_{i,L})=
v_f\sum_{i=1,2}\Psid_i\Psi_i,
\eqno (3.3)
$$
and
$$
\rho=\sum_{i=1,2}(\psid_{i,R}\psi_{i,R}+\psid_{i,L}\psi_{i,L})
=\sum_{i=1,2}\Psid_i\sigma_3\Psi_i.
\eqno (3.4)
$$

The quadratic form for the Fermi fields is approximated by
linearizing about the Fermi surface and replacing
$-\partial_x^2/2m-\mu$
 by $-iv_f\partial_x$. Thus the lagrangian density becomes
$$
\eqalign {L=\Psid_1\left(\matrix{\partial_\tau-iv_f\partial_x & \Delta\cr
        \Delta^*&\partial_\tau+iv_f\partial_x\cr}\right) \Psi_1 \cr
+ \Psid_2\left(\matrix{\partial_\tau-iv_f\partial_x & -\Delta\cr
        -\Delta^*&\partial_\tau+iv_f\partial_x\cr}\right) \Psi_2. \cr
}
\eqno (3.5)
$$
This lagrangian density is the same as that for a relativistic system,
but with $v_f$ replacing the speed of light.  The identities
for bosonizing  a relativistic system with the speed of light equal to
unity (see [13] for a review)
tell us that a Fermi action with a chiral mass term
$$
L_F=\psid(\gamma_\mu\partial_\mu+Me^{i\gamma_5\phi})\psi
\eqno (3.6)
$$
is equivalent (up to normal orderings) to the  bosonic action
$$
L_B=\frac 12 (\partial_\mu
\varphi^2)+M\cos(2\sqrt{\pi}\varphi-\phi).
\eqno (3.7)
$$

Provided the mass $M$ is large enough, the cosine term suppresses
fluctuations of the $\varphi$ field, forcing
$2\sqrt{\pi}\varphi\approx\phi$. With this relation
the Bosonic action  $L_B$ reduces to
$$
L_{eff}=\frac 1{8\pi} (\partial_\mu\phi^2).
\eqno (3.8)
$$
Corrections to $L_{eff}$ due to fluctuations in $\varphi$ will
contain higher gradients of $\phi$. On dimensional grounds
each $\partial_x$ will be accompanied by a  factor of $M^{-1}$.
Gradients of $M$ will be accompanied by a factor of $M^{-2}$.

Using these results, and restoring
$v_f$, we find that the real-time action becomes
$$
S_0= (2)\int dxdt\left\{ \frac{1}{8\pi v_f}\left
(\frac{\partial\phi}{\partial t}\right)^2
-\frac {v_f}{8\pi}\left(\frac{\partial\phi}{\partial
x}\right)^2\right\}.
\eqno (3.9)
$$
where $\Delta=|\Delta|e^{i\phi}$. The factor of  $(2)$ outside
the integral is from the two spin components.

 At the level of approximation we are using here (we have assumed that
the  gap is  small compared to the Fermi  energy, and that it
varies slowly in space and time) amplitude fluctuations of the gap do
not contribute.

How should we  extend (3.9) to three dimensions?  Let us first assume
$\Delta$ varies  in the $x$ direction only. The quadratic form  (2.3)
then decouples into a sum of one-dimensional lagrangians labled by the
transverse momenta $k_y$, $k_z$, so we have a set of  one-dimensional
problems to solve. Summing  the bosonized lagrangians of the form (3.9)
for each $k_y$, $k_z$ point on the Fermi surface gives us
$$
S=(2)L_yL_z\int \frac{d^2k_\perp}{(2\pi)^2}\int dxdt\left\{ \frac{1}{8\pi
v_f\cos\theta}\left
(\frac{\partial\phi}{\partial t}\right)^2
-\frac {v_f\cos\theta}{8\pi}\left(\frac{\partial\phi}{\partial
x}\right)^2\right\}.
\eqno (3.10)
$$
The lengths  $L_y,L_z$ define the spatial extent of the system in
the directions perpendicular to $x$, and are needed to determine the
density of allowed momenta in the $y,z$ directions. The angle $\theta$
is the angle between the normal to the (spherical) Fermi surface and
the $x$ direction.

We can write $d^2k_\perp$ in terms of the angle $\theta$ as
$$
\int\frac{d^2k_\perp}{(2\pi)^2}\ldots
= \frac{k_f^2}{(2\pi)}\int_0^1
\sin\theta d(\sin\theta)\ldots
\eqno (3.11)
$$
Thus
$$
\int
\frac{d^2k_\perp}{(2\pi)^2}\cos\theta=\frac 13\frac{k_f^2}{2\pi},
\eqno(3.12)
$$
and
$$
\int
\frac{d^2k_\perp}{(2\pi)^2}\frac{1}{\cos\theta}=\frac{k_f^2}{2\pi}.
\eqno (3.13)
$$

A three-dimensional equivalent of (3.9) is  therefore
$$
S=(2)\frac{k_f^2}{2\pi}\int d^3xdt\left\{ \frac{1}{8\pi v_f}\left
(\frac{\partial\phi}{\partial t}\right)^2
-\frac {v_f}{3}\frac{1}{8\pi}\left(\frac{\partial\phi}{\partial
x}\right)^2\right\}.
\eqno (3.14)
$$

As a check, notice that setting  $\phi=2(\delta k) x$ corresponds to
to giving every electron an extra $\delta k$ of momentum. In this case
second term in the integrand gives
$$
(2)\cdot\frac 1{(2\pi)^3} \cdot\frac 43 \pi k_f^3\cdot \frac 1 {2m} (\delta
k)^2
\eqno (3.15)
$$
and,  after identifying the product of the first three factors as the
(both spin components) number density $\rho_0$, we find an expression we
recognize as  the correct kinetic energy of the system.

The factor of $1/3$ in the second term of (3.15) yields
a wave velocity of $c_s=v_f/\sqrt{3}$. This  is equal to  the {\it
hydrodynamic\/} sound velocity for a three dimensional Fermi gas.  The
effect of the gap on the dynamics of the system is thus to
replace  Fermi-liquid zero-sound by conventional density  waves
which propagate at a velocity determined only by the mean density and
the bulk modulus.

A similar argument for (3.14),  but one  that maintains manifest rotational
invariance and allows for variations of the order paramter in all three
directions,
comes from observing that the principal effect of
spatial variations in $\phi$ occurs in the coupling of antipodal
points. We can thus  again decompose that action into a sum  of
actions over the fermi surface.   To get the correct measure, we
write
$$
\frac {d^3k}{(2\pi)^3}=k^2 dk
\frac {d^2\Omega} {(2\pi)^3} \approx \frac{k_f^2}{\pi}\frac
{dk}{2\pi} \frac{d^2\Omega}{4\pi}.
\eqno (3.16)
$$
which shows us how to express integrals over momentum space as
integrals over the solid angle $d^2\Omega$ on the Fermi surface,
together with an integral over $dk$ perpendicular to the  surface.

We now write the three dimensional bosonized action as a surface average
$$
S=\frac 12 (2)\frac{k_f^2}{\pi}\int \frac{d^2\Omega}{4\pi} \int
d^3xdt\left\{\frac 1{8\pi v_f}\left (\frac {\partial
\phi}{\partial t}\right)^2-\frac {v_f}{8\pi} ({\bf n}\cdot
\nabla\phi)^2\right\}.
\eqno (3.17)
$$
The symbol  ${\bf n}$ denotes a unit vector in real space
directed along the direction of $k$. The
initial $\frac 12$ arises  because  each ${\bf n}$   accounts for
a pair of antipodal  points on the Fermi sphere, and we do not
want to over-count.

Using
$$
\int \frac{d^2\Omega}{4\pi}n_in_j=\frac 13 \delta_{ij},
\eqno (3.18)
$$
we find
$$
S=(2)\frac {k_f^3}{24\pi^2}\frac 1 m
\int d^3x dt \left\{\frac 12 \frac
3{v_f^2}\left(\frac{\partial\phi}{\partial
t}\right)^2-\frac 12\left(\nabla \phi \right)^2\right\}.
\eqno (3.19)
$$

This action has appeared in the literature ({\it e.g.\/} [22]) but it
is not galilean invariant. It does not, without further assumptions,
reduce to conventional superfluid dynamics, and does not give rise to
the expected Magnus force on vortices. In the next section we will
exploit gauge-invariance and find what has gone wrong.

\line{\bf 4.) Gauge Invariance\hfil}

In relativistic field theory effective-action expansions are often
ambiguous up to polynomial counterterms.  These counterterms are
determined by imposing symmetry requirements such as gauge invariance.
Because of our use of a ``relativistic'' approximation, we must do this
here.  We  first use our  knowledge of the transformation
properties of $\Delta$  to find the effective interaction in the
presence of an external gauge  field. (This field is being used only as
a  probe, and its inclusion does not mean that we are considering a
charged superfluid.), and then fix the counterterms.

At a minimum, gauge-invariance requires us to replace plain derivatives
by covariant derivatives, so the effective action should contain the
terms
$$
S_0=\frac{\rho_0} {4m}
\int d^3x dt \left\{\frac 12 \frac
3{v_f^2}\left(\frac{\partial\phi}{\partial t} - 2eA_0\right)^2
-\frac 12\left(\partial_\mu \phi-2eA_\mu\right)^2\right\}.
\eqno (4.1)
$$

Differentiating with respect to the gauge field gives the
number current,
$$
j_\mu = \frac {\rho_0}{2m}
(\partial_\mu\phi-2eA_\mu),
\eqno (4.2)
$$
and  number density,
$$
\rho \buildrel ?\over = -\frac {\rho_0}{2mc_s^2}
\left(\frac{\partial\phi}{\partial t}-2eA_0\right).
\eqno (4.3)
$$

The first of  these expressions is the expected form of
the  current.  The second is consistent with the  number
conservation  law that follows from the equation of motion for
$\phi$, but it does not seem quite right. We expect  an
equilibrium density of fluid, $\rho_0$, to be present even when
the right hand side of (4.3) is zero. This  ``vacuum'' charge has been
omitted in (2.41). This should not have surprised us because  the
bosonization formul{\ae} give  bosonic expressions for
{\it normal-ordered\/} currents.

To get the correct number density   requires adding an $e\rho_0A_0$
counterterm. Maintaining gauge invariance further requires us to add a time
derivative  of $\phi$ so as  to complete the covariant derivative.
Thus we need  a term
$$
S_1=\int d^3xdt \frac 12
\rho_0\left(2eA_0-\frac{\partial\phi}{\partial t}\right).
\eqno (4.4)
$$
That the  time derivative is necessary is also indicated by its
occurence in (2.19). It appears naturally in the full diagramatic expansion
[14].

 Being a total derivative, the $\partial_t\phi$ term does not affect
the classical equations of motion, but it is  part of the key to
resolving the first-order versus second-order dynamics conundrum. To
see its significance consider its role in accounting for the Magnus
force. In the Matsubara formalism we sum over periodic configurations
--- but this does not imply that derivative $\partial_t \phi$
integrates to zero. Only the order parameter itself is required to be
periodic. The order parameter {\it phase\/} $\phi$ can be multivalued.
When there are vortices in the system every point that is encircled by
a vortex trajectory will have its value of $\phi$ incremented by
$2\pi$. The time derivative in (4.4) thus contributes a phase
proportional to the area enclosed by the trajectory. This phase
accounts for the  Magnus force on the vortex [9].  That this is so
should be clear by analogy with a particle moving in a magnetic field.
Here the action also accumulates a phase proportional the the area
enclosed by the particle trajectory. The Magnus force is the direct
analog of the Lorentz force on the particle. When the $\partial_t \phi$
term is dropped, as for example in [23], we loose the most important
part of the dynamics.

Our total action is now
$$
\eqalign{
S=&\int d^3xdt\bigg\{\frac 12
\frac{\rho_0}{mc^2_s}\left(\frac{\partial\phi/2}{\partial t}-eA_0\right)^2
-\frac {\rho_0}{2m}\left(\nabla\phi/2-e{\bf A}\right)^2\cr
&-\rho_o\left(\frac{\partial \phi/2}{\partial t}
-eA_0\right)\bigg\}.\cr
}
\eqno(4.5)
$$

We could rest content with (4.5) and extract
superfluid mechanics from it directly.  A more illuminating approach is to use
(4.5) to derive a Schr{\"o}dinger-like equation of motion for the
condensate.  To do this we  must promote the density  $\rho$ to the
status of a dynamical variable.

If we write
$$
S=\int d^3xdt
\left\{- \rho \left(\frac{\partial\phi/2}{\partial t}-eA_0\right)
-\frac {\rho_0}{2m}\left( \nabla \phi/2-e{\bf A}\right)^2
-\frac 12 \frac{c_s^2 m}{\rho_0}(\rho-\rho_0)^2\right\},
\eqno (4.6)
$$
then integrating out the $\rho$ field, or eliminating it by using its
equation of motion, gives (4.5). To obtain a galilean invariant
expression we must also replace the $\rho_0$ in front of the second
term in the integrand of (4.5) by the full density, $\rho$. This step
takes us beyond the approximations we have been using, but is justified
because we know that the exact result must be consistent with (2.18)
which requires the coefficient to be $\rho$, and not $\rho_0$. Notice
also that omiting the $\rho_0 A_0$ counterterm would result in  the
$(\rho-\rho_0)^2$ in the last  term being replaced by plain $\rho^2$.

Now {\it define\/} a field $\Psi=\sqrt{\frac \rho 2}e^{i\phi}$.   For mnemonic
purposes $\Psi$  can  be thought of as a ``wavefunction'' of the Cooper
pairs.  It must {\it not}  however  be confused with the BCS order
parameter $\Delta$. There is in general no simple relation between
$|\Delta|$ and $\sqrt{\rho}$. We also define the symbol $e^*=2e$ which
can be though of as the Cooper-pair charge, and the symbol $m^*=2m$
which can be though of as  the Cooper-pair mass.  Finally we set
$\rho_0^*=\rho_0/2$,  and this can be though of as the Cooper-pair
density.

With these definitions we can write (4.6) in a appealingly simply
form.  Up to higher-order gradients of $\rho^*$, it is equivalent  to
$$
S=\int d^xdt\left\{i\Psid(\partial_t-ie^*A_0)\Psi-\frac
1{2m^*}|(\nabla-ie^*{\bf A})\Psi|^2-\frac \lambda 2
(|\Psi|^2-\rho_0)^2\right\},
\eqno (4.7)
$$
where $\lambda={c_s^2 m^*}/{\rho_{0}^*}$.

Varying this last action
gives rise  the
galilean-invariant  Gross-Pitaevskii non-linear Schr{\"o}dinger equation for
$\Psi$
$$
i(\partial_t-ie^*A_0)\Psi=-\frac
1{2m^*}(\nabla-ie^*{\bf A})^2\Psi
+\lambda(|\Psi|^2-\rho_0^*)\Psi.
\eqno (4.8)
$$
It can be rewritten  as the Euler equation for a compressible fluid
(See appendix), and automatically yields the correct vortex dynamics.

\line{\bf Discussion\hfil}

 While it is well known that topological terms in an action can affect
the quantum mechanical properties of the  system, it is at first sight
surprising that the total derivative  $\partial_t \phi$ in the
counterterm  (4.4) can affect  the classsical equations of motion.
 Nevertheless omiting  (4.4) deletes the $\rho_0$ from (4.6) and (4.8)
and so sets the equilibrium value of  $\langle \Psi\rangle $ to zero.
This alteration starts out being largely cosmetic.  Once we  express
$\Psi$ in terms of $\rho$ and $\phi$, and  eliminate $\rho$, the
equation of motion for $\phi$ remains unaffected.  What is really lost
is the interpretation of  the auxiliary variable $\rho$ as the fluid
density.  This interpretation  is  however essential in justifying our
replacement of $\rho_0$ in the coefficient of the kinetic energy term
by $\rho$. The $\rho$ substitution in turn modifies (4.4) by adding
the Bernoulli effect pressure change to the potential $A_0$, and is
necessary for rewriting everything in terms of $\Psi$. The total
derivative thus has its effect indirectly by encouraging other
modifications.

The Gross-Pitaevskii equation  (4.8) does not have the wide range of
applicability of its Bose superfluid cousin. In the Bose superfluid the
Gross-Pitaevskii equation  gives qualitatively good results even when
the density varies rapidly  and widely from its equilibrim value.  Our
derivation holds only for small and slow deviations. Applying it to a
vortex where $|\Psi|$ vanishes  might be therefore be questioned ---
but the identification of the Magnus force with a Berry phase [9] shows
that the vortex motion is determined by topological effects arising in
the behaviour of the fluid away from the singular core.  The
Gross-Pitaevskii equation, with a non-zero $\langle \Psi \rangle$, accommodates
these effects and remains useful  therefore  as an
 insight-providing  first approximation to the dynamics.

The quantity $\Psi$ is neither the BCS order parameter, nor a genuine
Cooper-pair wavefunction. It is simply a mathematical construct that
permits us to write the low-energy effective action  in an easily
digestible form. If a similar variable had been introduced in reference
[4], for example; then the conclusions of that work would be seen
to be compatable with the ones presented here.  Because of the
artificial variable $\Psi$, the Gross-Pitaevskii equation (4.8) should
not be thought of as a ``time-dependent Landau-Ginzburg equation''. A
true Landau-Ginzburg equation would contain information about the
dynamics of the order parameter amplitude. This amplitude mode has a
frequency gap, so it does not appear in the low-energy description.

{\it Note Added\/}: While this manuscript was in preparation I received a copy
of
[25] which contains conclusions similar to those presented here.

\line{\bf Acknowledgements\hfil}

This work was begun at the Institute for Theoretical Physics in Santa
Barbara, and  was  supported by the National Science Foundation under
grant numbers PHY89-04035 and DMR91-22385.  I must thank Ping Ao for
many conversations  and for sending me copies of his work before
publication.  I would like  thank Daniel Boyanovsky, and Tony Zee  for their
comments, and also Paul Goldbart whose attentive reading of an earlier
version of this manuscript has lead to many improvements.

\vskip 20 pt
\line{\bf Appendix. Madelung Fluids.\hfil}

In this appendix we  review the fluid dynamic interpretation of
(4.8). In order to distinguish quantum from classical effects we will
include explicit factors of $\hbar$.

Given a time-dependent non-linear Schr{\"o}dinger equation  of
the form
$$
i\hbar(\partial_t-ieA_0/\hbar)\Psi=-\frac
{\hbar^2}{2m}\sum_{a=1}^{3}(\partial_a-ieA_a/\hbar)^2\Psi
+\lambda (|\Psi|^2-\rho_0)\Psi,
\eqno (A.1)
$$
we can recast it  as the equation of motion of a charged compressible
fluid. This observation was originally made by Madelung
(although without the non-linear term) very soon after the
discovery of the Schr{\"o}dinger equation [24].

We set $\Psi=\sqrt{\rho}e^{i\theta}$  and
define a velocity field ${\bf v}$ in such a way that the number-current
$$
{\bf j}=\frac {\hbar}{2mi}\left(\Psi^*(\nabla -ie{\bf A}/\hbar)
\Psi-((\nabla+ie{\bf A}/\hbar)\Psi^*)\Psi\right)
\eqno(A.2)
$$
may be written ${\bf j}=\rho{\bf v}$. This leads to
$$
{\bf v}=\frac \hbar m (\nabla\theta-e{\bf A}/\hbar).
\eqno(A.3)
$$

In the absence of vortex singularities in $\Psi$, the  vorticity,
$\omega=\nabla\wedge {\bf v}$, is completely determined by the
gauge field to be
$
\omega=-\frac em \nabla\wedge {\bf A}=-\frac{e{\bf B}}{m},
$
{\it i.e}
$$
m\omega+e{\bf B}=0.
\eqno(A.4)
$$
When the gauge field is dynamical, and not just an external
probe, this equation is responsible for the Meissner effect. A
penetrating ${\bf B}$ field implies a  uniform vorticity which
would lead, in a  sphere of radius $R$,  to a kinetic energy  that
grows as $R^5$, {\it i.e} faster than extensive. More precisely,
 taking the curl of the equation
$
{\nabla \wedge} {\bf B}=e{\bf j}$, and using  ${\bf j}=\rho {\bf v}_s$ implies
that
$$
\nabla^2 {\bf B}-\frac {e^2\rho}{ m }{\bf B}=0,
\eqno(A.5)
$$
which leads to flux screening.

With the definition(A.3) the imaginary  and real parts of (A.1)
become respectively the continuity equation
$$
\partial_t\rho+\nabla\cdot \rho{\bf v}=0,
\eqno(A.6)
$$
and the Euler equation  governing the flow of a barotropic fluid
$$
m(\partial_t {\bf v}+{\bf v}\cdot \nabla {\bf v})=
e({\bf E}+{\bf v}\wedge {\bf B})-\nabla \mu.
\eqno(A.7)
$$
The word {\it barotropic\/}  refers to the
simplifying property that the pressure term $\frac1\rho\nabla P$
which occurs on the right hand side of the conventional Euler
equation is here combined into the gradient of a potential
$$
\mu=\lambda(\rho-\rho_0)-\frac {\hbar^2}{2m}\frac{\nabla^2 \rho}{\rho}.
\eqno(A.8)
$$
The potential contains the expected compressibility pressure, depending
on the deviation from the equilibrium density, plus a correction
depending on gradients of $\rho$. This correction is called the {\it
quantum pressure}.  Notice that $\hbar $ does not appear in (A.4 - 7),
except implicitly in the quantum pressure.
 For small density variations, and those are the only ones for which
our derivation of the Gross-Pitaevskii equation is valid, the quantum
pressure term is unimportant.

The Euler equation (A.7) is   derived by first taking the gradient of
(A.1)  and finding the equivalent Bernoulli form
$$
m(\partial_t {\bf v} -{\bf v}\wedge \omega)= e({\bf E}+{\bf v}\wedge{\bf
B})-\nabla\left(\frac 12 m {\bf v}^2+\mu\right).
\eqno(A.9)
$$
A cancellation  of the $m{\bf v}\wedge \omega$ term against
the  $e{\bf v}\wedge{\bf B}$  term is evident on use of (A.4).
It is after this cancellation, and so  without reference to ${\bf
B}$ or $\omega$, that the hydrodynamic picture of
superconductivity is conventionally displayed  [5]. I prefer to
keep $\omega$ and $B$   in (A.9) and rewrite it as (A.7).
Then one can see that the only difference between  the superfluid
dynamics of the condensate and  ordinary fluid dynamics lies in
the constraint (A.4).

\line{\bf References\hfil}

\item{[1]}  L.~P.~Pitaevskii,  Zh. Eksp. Teor. Fiz. {\bf 40}
(1961) 646, Translated in Sov. Phys. JETP {\bf 13} (1961) 451.

\item{[2]} E.~P.~Gross, Nuovo Cimento {\bf
20} (1961) 454; J. Math Phys. {\bf 4} (1963) 195.

\item{[3]} R.~P.~Feynman, {\it Statistical Mechanics\/}, (Benjamin 1972).

\item{[4]} E.~Abrahams, T.~Tsuneto, Phys. Rev  {\bf 152} (1966)
152.

\item{[5]} W.~F.~Vinen, {\it A Comparison of the properties of
superconductors and superfluid helium}, in {\it
Superconductivity}, R.~D.~Parks ed., (Marcel Decker NY 1969); See also
M.~Cyrot, Rep. Prog. Phys {\bf 36} (1973) 103.

\item {[6]} M.~Dreschler, W.~Zwerger, Annalen der Physik, {\bf 1}, (1992) 15;
M.~Randeria, \hfil\break C.~A.~R.~Sa~de~Melo, J.~R.~Engelbrecht,
Physica B  {\bf 194-196} (1994) 1409.

\item{[7]} H.~T.~C.~Stoof, Phys. Rev. {\bf 47} (1993) 7979.

\item{[8]} P.~Ao, D.~L.~Thouless, X-M.~Zhu, unpublished.

\item{[9]} P.~Ao, D.~L.~Thouless,
   Phys. Rev. Lett.,  {\bf 70} (1993) 2158.

\item{[10]} J.~C.~Neu,  Physica D{\bf 43} (1990) 407.

\item{[11]} J.~C.~Neu,  Physica D{\bf 43} (1990) 385.

\item{[12]} P. Nozi\`ers,   W. F. Vinen, Phil. Mag. 14 (1966) 667.

\item{[13]} {\it Bosonization}, M.~Stone ed. (World Scientific 1994).

\item{[14]} A.~M.~J. Schakel, Mod. Phys. Lett. B {\bf 4} (1990) 927.

\item{[15]} A.~Luther, Phys. Rev {\bf B19} (1979) 320.

\item{[16]} F.~D.~M.~ Haldane, unpublished.

\item{[17]} A.~Houghton,
J.~B.~Marston, Phys. Rev. {\bf B48} (1993) 7790.

\item{[18]} A~.H.~Castro Neto, E.~Fradkin, Phys. Rev. Lett.
{\bf 72} (1994) 1393.

\item{[19]} P.~W.~Anderson,  Phys. Rev.
Lett. {\bf 64} (1990) 1839.

\item{[21]} F.~Gaitan, M.~Stone, Annals of Phys. (NY) {\bf
178} (1987) 89.

\item{[22]} See N.~R.~Werthamer, {\it The Ginzburg Landau
equations and their extensions}, in {\it Superconductivity},
R.~D.~Parks ed., (Marcel Dekker NY 1969).

\item{[23]}  V.~N.~Popov, V. N.
 {\it  Functional integrals and collective excitations\/}, (Cambridge
    University Press,1987).

\item{[24]} E.~Madelung, Z. Phys {\bf 40} (1927) 322.

\item{[25]} I.~J.~R.~Aitchison,  P.~Ao, D.~L.~Thouless, X-M.~Zhu, {\it
Effective
Theories of BCS Superconductors at $T=0$\/}. CERN preprint TH-7385/94.

\end